\documentclass[9pt,twocolumn,twoside]{opticajnl}
\journal{opticajournal} 
\pdfoutput=1
\setboolean{shortarticle}{false}

\usepackage{lineno}
\usepackage{caption}
\title{In-situ tuning of optomechanical crystals with nano-oxidation}

\author[1,2,*]{Utku Hatipoglu}
\author[1,2,*]{Sameer Sonar}
\author[1,2]{David P. Lake}
\author[1,2]{Srujan Meesala}
\author[1,2,3,$\dagger$]{Oskar Painter}

\affil[1]{Institute for Quantum Information and Matter, California Institute of Technology, Pasadena, CA 91125, USA}
\affil[2]{Kavli Nanoscience Institute and Thomas J. Watson, Sr., Laboratory of Applied Physics, California Institute of Technology, Pasadena, CA 91125, USA}
\affil[3]{AWS Center for Quantum Computing, Pasadena, CA 91125,
USA}

\affil[*]{These authors contributed equally to this work.}
\affil[$\dagger$]{opainter@caltech.edu}

\begin{abstract}
Optomechanical crystals are a promising device platform for quantum transduction and sensing. Precise targeting of the optical and acoustic resonance frequencies of these devices is crucial for future advances on these fronts. However, fabrication disorder in these wavelength-scale nanoscale devices typically leads to inhomogeneous resonance frequencies. Here we achieve in-situ, selective frequency tuning of optical and acoustic resonances in silicon optomechanical crystals via electric field-induced nano-oxidation using an atomic-force microscope. Our method can achieve a tuning range $>2$~nm ($0.13\%$) for the optical resonance wavelength in the telecom C-band, and $>60$~MHz ($1.2\%$) for the acoustic resonance frequency at $5$~GHz. The tuning resolution of $1.1$~pm for the optical wavelength, and $150$~kHz for the acoustic frequency allows us to spectrally align multiple optomechanical crystal resonators using optimal oxidation patterns determined via an inverse design protocol. Our results establish a method for precise post-fabrication tuning of optomechancical crystals. This technique can enable coupled optomechanical resonator arrays, scalable resonant optomechanical circuits, and frequency matching of microwave-optical quantum transducers.
\end{abstract}

\setboolean{displaycopyright}{false} 

\begin{document}

\maketitle

\section{Introduction}
 
 Optomechanical crystals (OMCs) provide a coherent interface between optical photons and acoustic phonons \cite{maccabe2020nano}. This capability is now being utilized in microwave-optical quantum transducers \cite{mirhosseini2020superconducting, meesala2023non, jiang2023optically, weaver2023integrated,zhao2023electro} towards connecting gigahertz frequency superconducting quantum processors \cite{Kjaergaard2020} via low-loss optical communication channels. In parallel efforts, OMCs have enabled coherent control and routing of phonons in chip-scale optomechanical circuits \cite{fang2016optical,Patel2018, Zivari2022} with demonstrations of nonreciprocal optical transmission \cite{fang2017generalized} and topologically protected phonon transport \cite{ren2022topological}. However, scaling to multi-node quantum networks and more complex optomechanical circuits is challenging due to variations in optical and acoustic resonance frequencies across devices caused by fabrication imperfections. In particular, since critical feature sizes in OMCs are affected by the precision limits of electron beam lithography and reactive ion etching, the spread in resonance frequencies across devices can be more than a hundred times the resonance linewidths, thereby hindering frequency alignment. 
 
\begin{figure*}[t]
\includegraphics[width=17.1cm]{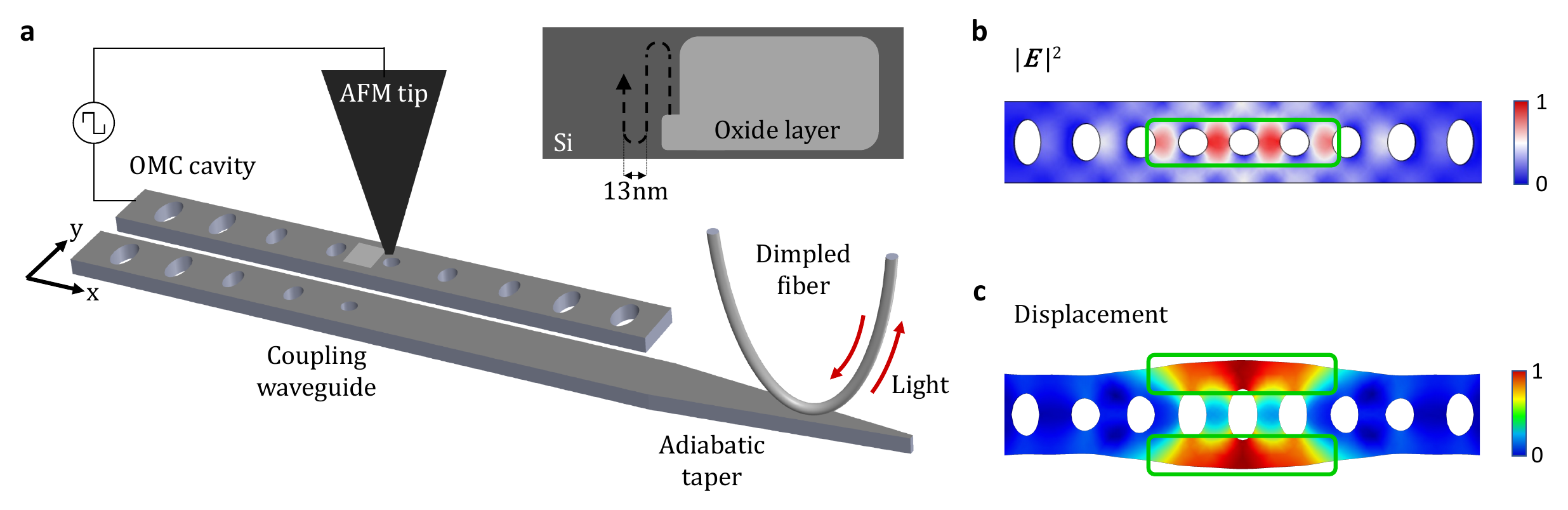}
\centering
\caption{\textbf{Nano-oxidation setup schematic and OMC cavity mode profiles.} \textbf{a}, Simplified schematic of the AFM nano-oxidation setup. The AFM tip is used to perform nano-oxidation while measuring the optical and acoustic resonances of the device in real-time with a dimpled optical fiber. The AFM tip is operated in tapping mode over a grounded silicon sample and biased with a square wave voltage, allowing for electrochemical formation of an oxide layer on the silicon surface. Inset illustrates the raster scan used to generate the oxide layer. \textbf{b}, Electric field intensity profile of the optical mode. The optics-focused oxidation region is outlined in green. \textbf{c}, Displacement magnitude profile of the acoustic breathing mode. The acoustics-focused oxidation regions are outlined in green.}
\label{fig:fig1}
\end{figure*}

\begin{figure}[h!]
\includegraphics[width=9cm]{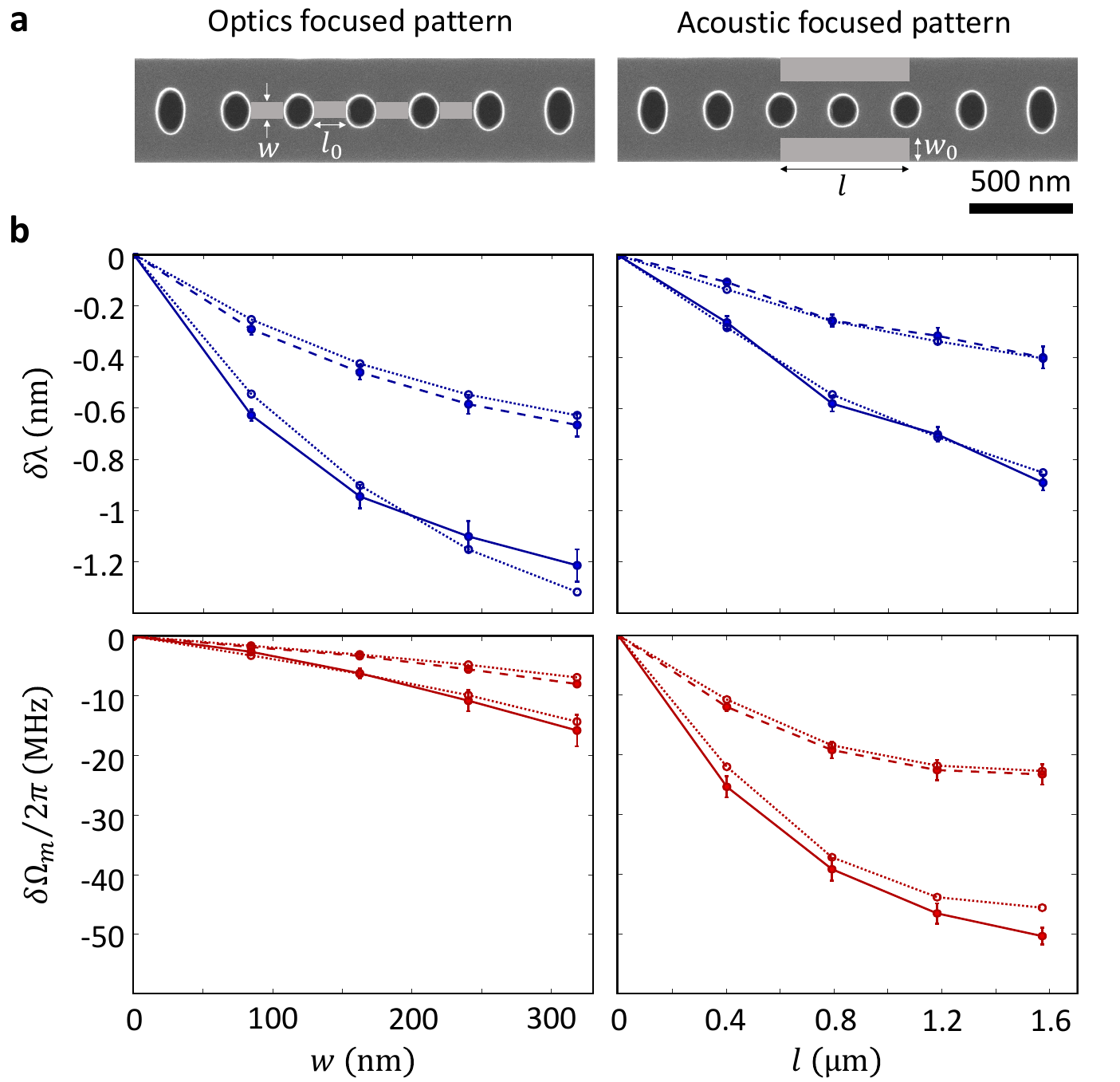}
\centering
\caption{\textbf{Coarse frequency tuning of OMC cavity.} \textbf{a}, Oxidation patterns used for coarse tuning of optical (left) and acoustic (right) resonances. For the optics-focused pattern, the length is fixed at $l_0=170$ nm and the width, $w$ is varied. For the acoustics-focused pattern, the width is fixed at $w_0=120$ nm and the length, $l$ is varied. \textbf{b}, Tuning of optical and acoustic resonances as a function of the size of the oxidation pattern. Optical wavelength shift $\delta\lambda$ with respect to $w$ (top left) and $l$ (top right). Acoustic frequency shift $\delta\Omega/2\pi$ with respect to $w$ (bottom left) and $l$ (bottom right). Solid (dashed) lines represent data taken for aggressive (mild) oxidation. Dotted lines show simulated values from finite element method simulations. Error bars represent the standard deviation across five OMC cavities measured for each dataset.}
\label{fig:fig2}
\end{figure}

Post-fabrication tuning of optical resonance frequencies in chip-scale microcavities has been achieved using a variety of techniques including laser-assisted thermal oxidation \cite{panuski2022laser}, atomic force microscope (AFM) nano-oxidation \cite{hennessy2006tuning,yokoo2011ultrahigh}, strain tuning \cite{wong2004strain,luxmoore2012restoring}, thermo-optic tuning \cite{nawrocka2006tunable,faraon2007local}, and gas condensation \cite{mosor2005scanning, strauf2006frequency}. However, since OMCs co-localize optical and acoustic resonances in a wavelength-scale volume, selective tuning of both optical and acoustic resonances without compromising the quality factors of either is a significantly more complex endeavor and an outstanding technical challenge. Here we use AFM nano-oxidation tuning to demonstrate such control over the resonance frequencies of OMCs. In our approach, field-induced oxidation of the silicon device surface with high spatial resolution allows us to tune the acoustic resonance by modifying the local mass distribution and elasticity, and the optical resonance by modifying the local refractive index. By using an inverse design protocol to guide the nano-oxidation sequence in real-time, we achieve simultaneous alignment of the optical and acoustic resonance frequencies of multiple OMC cavities.

\section{AFM nano-oxidation setup}
A simplified schematic of the experimental setup is shown in Fig.\,\ref{fig:fig1}a. We perform nano-oxidation using a conductive chromium/platinum coated silicon AFM tip with a radius <25 nm \cite{parksystems}, and track the optical and acoustic resonances in real-time by performing optomechanical spectroscopy via a dimpled optical fiber coupled to the device. The AFM is operated in tapping mode while a voltage bias is applied to the conductive tip, and the silicon-on-insulator (SOI) chip with a silicon device layer resistivity of 5 k$\Omega$-cm is grounded. When the voltage-biased AFM tip is brought close to the silicon surface, a strong electric field triggers an electrochemical reaction between ions in the native water meniscus and the silicon surface, resulting in local oxidation of silicon. Importantly, this reaction can proceed even when the silicon surface is covered with native oxide, because the strong local electric field allows oxyanions (OH$^{-}$ , O$^{-}$) to diffuse through the native oxide \cite{garcia2006nano}. In our experiments, we maintain the AFM setup enclosure at ambient conditions with a relative humidity of 40$\pm$5$\%$ and temperature of 23$\pm$1$^\circ$C, and apply a square wave voltage oscillating between $\pm 10$ V at a frequency of 20 Hz to the AFM tip.

\begin{figure*}[t]
\includegraphics[width=\textwidth]{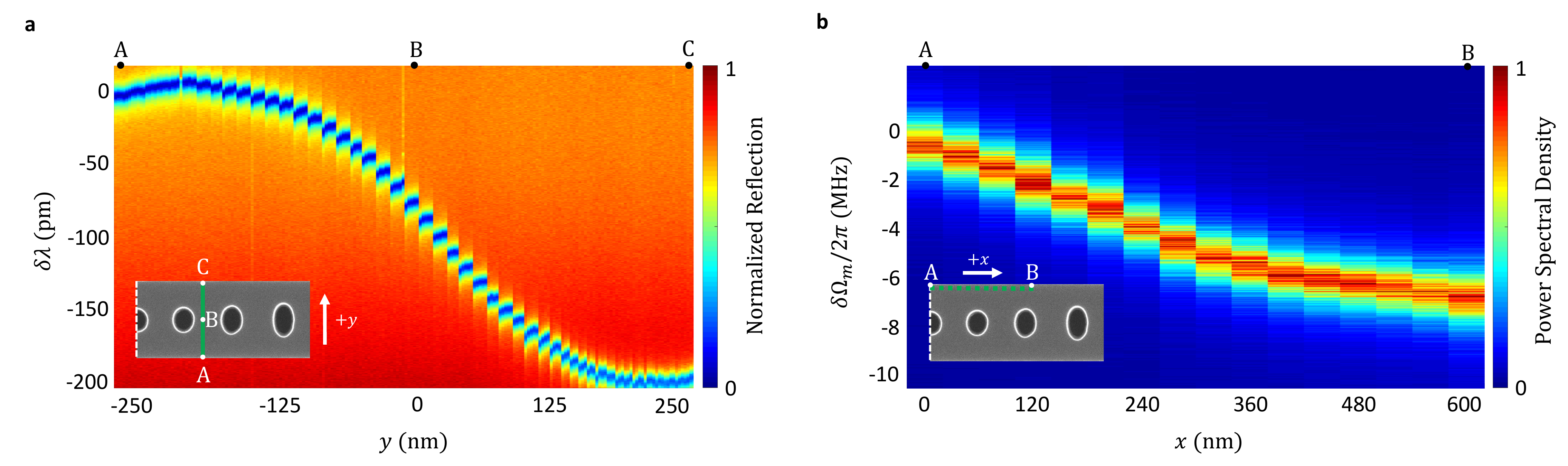}
\centering
\caption{\textbf{Real-time, in-situ monitoring of the oxidation process.} \textbf{a}, Real-time spectra showing optical resonance tuning during sequential application of the oxidation pattern. As shown in the inset, the pattern is applied starting at point A, one pixel at a time, up to point C. AFM tip locations during the experiment are shown at the top of the plot with letters A, B and C which correspond to the points labeled on the inset image. Discontinuities along the horizontal axis are due to proximity effects between the AFM tip and the OMC cavity during the oxidation sequence. \textbf{b}, Real-time data showing acoustic frequency shift during pixel-wise oxidation. Oxide pixels are applied on the surface starting from point A up to point B as shown in the inset.}
\label{fig:fig3}
\end{figure*}

The OMC cavities in this work were fabricated on an SOI chip with a $220$~nm thick silicon device layer. The cavities were patterned via electron beam lithography followed by reactive ion etching, and finally suspended by removing the buried oxide layer with an anhydrous vapor hydrofluoric acid etch. The devices were designed to support a fundamental TE-like optical resonance at a wavelength of $\sim$1550 nm and a breathing acoustic resonance at a frequency of $\sim$5.1 GHz (Fig.\,\ref{fig:fig1}b,c) \cite{OMCDesign}. An on-chip adiabatic waveguide coupler \cite{groblacher2013highly} allows for optical coupling to the OMC cavity using a dimpled optical fiber. The optical spectrum of the OMC cavity was probed in reflection mode, and the resonance frequencies were recorded before and after each step of the nano-oxidation sequence. To measure the acoustic spectrum of the device, we routed the optical signal reflected from the device to a high-speed photodetector and measured the power spectral density (PSD) of the photodetector electrical output on a spectrum analyzer. The result is proportional to the PSD of thermal displacement fluctuations of the acoustic mode which are transduced onto the optical signal via the optomechanical interaction in the device. In these measurements, the laser was blue-detuned with respect to the optical resonance frequency by a detuning close to the acoustic resonance frequency, and operated at low power to minimize optomechanical back-action on the acoustic mode \cite{aspelmeyer2014cavity}.

\section{Nano-oxidation characterization}

From the simulated mode profiles shown in Fig.\,\ref{fig:fig1}b,c, we identified two strategic oxidation regions for selective tuning of the optical and acoustic modes. Nano-oxidation in the region shown with the green rectangle in Fig.\,\ref{fig:fig1}b is expected to induce a relatively large change in the optical resonance due to a high concentration of electric field energy. Here the impact on the acoustic resonance is expected to be minimal due to two reasons. First, the displacement amplitude is small in this region. Second, even though the stress amplitude is large in this region, the Young’s modulus of the oxide layer when weighted by its thickness is similar to that of the original silicon. In contrast, nano-oxdiation on the regions shown with the green rectangles in Fig.\,\ref{fig:fig1}c leads to a large acoustic resonance shift and a small optical wavelength shift due to the high concentration of motional mass and low concentration of electric field energy. In section \ref{Supplementary_D} of the supplementary information, we show the results of finite element method (FEM) simulations where we investigated the optical and acoustic frequency shifts due to creation of an oxide pixel at an arbitrary location on the OMC surface. This allowed us to identify candidate regions for fine tuning of optical and acoustic resonance frequencies with maximal selectivity. For coarse frequency tuning, we opted to use larger, rectangular oxide patterns instead of oxide pixels to maximize the tuning range at the expense of selectivity.

Prior to experiments on OMC devices, we characterized the effect of the AFM tapping amplitude and scanning velocity on nano-oxidation using a bare silicon chip. We observed that lower tapping amplitude and slower scanning velocities result in wider and thicker oxide lines. We measured the three-dimensional profiles of field-induced oxide lines and pixels using AFM measurements as described in section \ref{supplementary_characterization} of the supplementary information. Based on the amount of frequency tuning required, we operated the AFM in two distinct modes which we refer to as `mild' and `aggressive' tuning modes. In the mild tuning mode, the tapping amplitude and scan speed were set to  15 nm and 100 nm/s, respectively, resulting in oxide thickness of approximately 1.2 nm. In the aggressive tuning mode, the tapping amplitude and scan speed were set to 5 nm and 50 nm/s, respectively, leading to oxide thickness of approximately 2.5 nm. In the mild tuning mode, nano-oxidation was found to generate single pixels with a lateral size of approximately 25 nm. In order to obtain thicker and more uniform oxide patterns with a raster scan, we used a line spacing of 13 nm with significant overlap between neighboring lines as illustrated in the inset of Fig.~\ref{fig:fig1}a. This procedure allowed us to generate oxide layers with a thickness of 1.6 nm and 3.2 nm in the mild and aggressive tuning modes, respectively.

\setlength{\parskip}{1pt}

After establishing these nano-oxidation parameters, we performed optical and acoustic resonance frequency tuning experiments on a single OMC cavity. We specifically studied the impact of two oxidation patterns: (i) an optics-focused pattern (Fig.\,\ref{fig:fig2}a left) composed of four rectangles located between the five central ellipses in the beam; here the lengths of the rectangles were kept constant at $l_0$ = 170 nm, while their width, $w$ was varied, and (ii) an acoustics-focused pattern (Fig.\,\ref{fig:fig2}a, right) composed of two rectangles situated at the edge of the OMC cavity; here the widths of the rectangles were kept fixed at $w_0 =$ 120 nm, while their length, $l$ was varied. As shown in Fig.\,\ref{fig:fig2}b, the experimentally measured tuning of the resonance frequencies (solid and dashed lines) agrees with simulation results (dotted lines). For the optics-focused pattern, increasing $w$ leads to a large reduction in the optical wavelength and a smaller reduction in acoustic frequency. For the acoustics-focused pattern, increasing $l$ leads to a decrease in acoustic frequency due to increasing motional mass ($\Omega_m \propto \sqrt{1/m}$) and a decrease in the optical wavelength due to a decrease in the effective dielectric constant. Aggressive oxidation at $w =$ 320 nm resulted in more than 1.2 nm blue shift of the optical wavelength while aggressive oxidation at $l$ = 1.6 $\mu$m resulted in more than 50 MHz reduction in the acoustic frequency. In Fig.~\ref{fig:fig2}b, each data point represents the average value and the standard deviation obtained from five devices. For each device, the patterns were applied cumulatively, starting from the center and adding more oxide to increase either $w$ or $l$ at each step. 

\begin{figure*}[t]
\includegraphics[width=\textwidth]{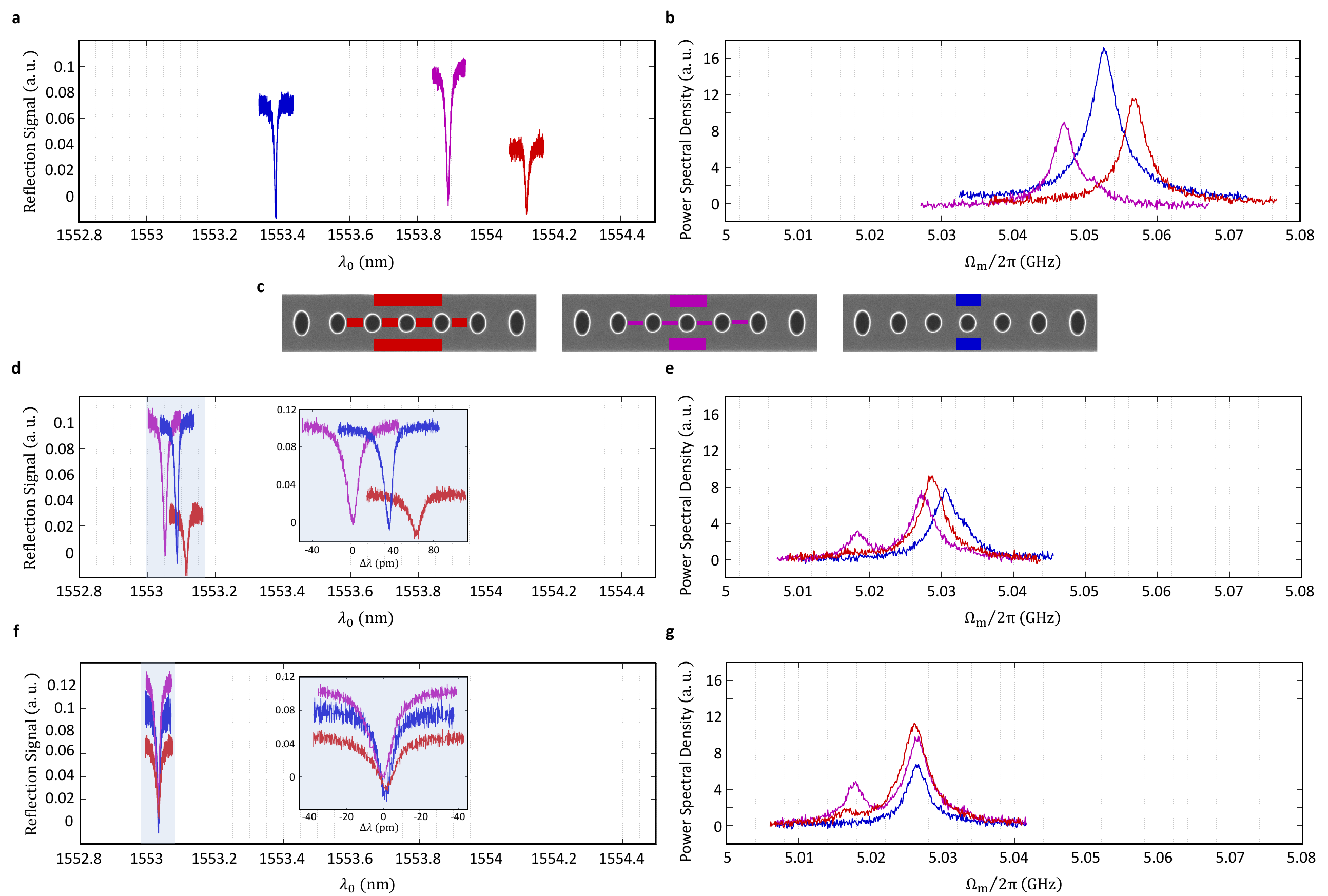}
\centering
\caption{\textbf{Simultaneous optical and acoustic resonance frequency alignment of three OMC cavities.} Initial \textbf{a}, optical and \textbf{b}, acoustic resonance spectra. \textbf{c}, Joint patterns used for coarse alignment of the three OMC cavities. The patterns are shown with a color code matched to the OMC spectra, and are applied in the order shown from left to right. Resonance frequencies were recorded after every step of the nano-oxidation protocol. \textbf{d}, Optical and \textbf{e}, acoustic spectra measured after completion of the coarse alignment of sequence. \textbf{f}, Optical and \textbf{g}, acoustic spectra after completion of the real-time fine-tuning sequence. Insets in panels d and f show a zoomed-in view of the optical resonances. The satellite acoustic resonance in magenta in panels e and g is due to a neighboring OMC cavity within the evanescent field of the optical coupling waveguide.}
\label{fig:fig4}
\end{figure*}

\section{Simultaneous optical and acoustic frequency tuning}

To achieve simultaneous tuning of optical and acoustic resonances, we define a joint pattern consisting of a union of the previously introduced optics-focused and acoustics-focused patterns. This is parameterized by geometric parameters $w$ and $l$, and binary variables $M_\text{O}$ ($M_\text{A}$) which denote mild or aggressive oxidation in the optical (acoustic) components of the joint pattern. Using the experimental data shown in Fig.\,\ref{fig:fig2}, we interpolate $\delta \lambda$ and $\delta \Omega_\text{m}$ as a function of the parameters $\{w,l,M_\text{O},M_\text{A}\}$. To achieve the desired frequency shifts, we developed an inverse design algorithm to find an optimal combination of the acoustic and optics focused patterns. The algorithm takes the desired frequency and wavelength shift as the input and generates the required joint oxidation pattern by searching for the most suitable combination of $l$ and $w$. The algorithm starts the search from milder oxidation modes to prevent a reduction in optical Q-factor as explained in supplementary section \ref{supplement:inverse_algo}.

While the inverse design algorithm is sufficient for coarse tuning of the OMC wavelength and frequency, fine frequency alignment to within the linewidth of the optical and acoustic resonances requires pixel-by-pixel nano-oxidation with real-time feedback. This method allows for $\sim$1 pm precision for optical wavelength tuning and $\sim$150~kHz precision for acoustic resonance tuning. To quantify the sensitivity of the optical resonance wavelength to single oxide pixels, we applied a pixel-by-pixel linear oxide pattern perpendicular to the OMC long axis with a pitch of 13 nm shown in the inset of Fig~\ref{fig:fig3}a. The resonance spectra recorded in Fig.\,\ref{fig:fig3}a for each incremental pixel in the oxidation sequence reveal a decrease in the per-pixel wavelength shift towards the edges of the OMC, with an observed shift of 1.1~pm/px at the extreme points A and C. This spatial dependence is in agreement with predictions from FEM simulations, as discussed in supplementary section \ref{Supplementary_D}. We perform a similar sensitivity analysis for the acoustic resonance via a pixel-by-pixel linear oxide pattern applied with a pitch of 40 nm along the edge of the OMC, shown in the inset of Fig~\ref{fig:fig3}b. The acoustic resonance spectrum is recorded after the addition of every oxide pixel, and the results are shown in Fig.\,\ref{fig:fig3}b. We measure a maximum frequency shift of -1 MHz/px at point A which is in line with the center of the OMC cavity. As we move away from the center, the sensitivity eventually decreases to a value of -150~kHz/px as measured at the extreme point B. 

After characterizing coarse and fine tuning of optical and acoustic resonance frequencies of a single OMC cavity, we used these techniques to perform frequency alignment of both the optical and acoustic resonances of three OMC cavities. Due to constraints on the maximum achievable tuning range, we pre-select cavities with initial optical wavelength spread <1 nm and acoustic frequency spread <20 MHz, shown on Fig.\,\ref{fig:fig4}a, b. We begin with coarse tuning steps using oxidation patterns generated by the inverse design algorithm as shown in Fig.\,\ref{fig:fig4}c. This aligns the optical resonance wavelengths across all cavities to within a tolerance of <200 pm and acoustic resonance frequencies to withina tolerance of 5 MHz. The results of the coarse tuning are shown on Fig.\,\ref{fig:fig4}d,e. We then proceed to real-time, pixel-by-pixel oxidation with feedback to achieve fine tuning of the resonance frequencies. After four rounds of fine-tuning, the optical and acoustic resonance frequencies were successfully tuned to within 2 pm and 500 kHz respectively (see Fig~\ref{fig:fig4}f, g). A step-by-step description of the frequency alignment procedure is described in detail in supplementary section \ref{3_OMC_tuning_supplement}.

\section{Outlook}

We have shown simultaneous, real-time tuning of optical and acoustic resonances in OMC cavities via AFM nano-oxidation, demonstrating optical wavelength and acoustic frequency shifts >2 nm (0.13\%) and >60 MHz (1.2\%), respectively. Such in-situ tuning methods could enable the realization of coupled optomechanical resonator arrays to study topological phases of photons and phonons \cite{peano2015topological,peano2016topological,brendel2017pseudomagnetic,ren2022topological}. In the context of OMC-based microwave-optical quantum transducers \cite{mirhosseini2020superconducting, meesala2023non, jiang2023optically, weaver2023integrated,zhao2023electro}, the nano-oxidation tuning demonstrated here will enable post-fabrication frequency alignment of transducers in multiple nodes of a quantum network. Moreover, these techniques are applicable to other material systems such as silicon nitride \cite{chien2001nano}, where tuning of optomechanical devices may be required \cite{delaney2022}. Finally, AFM nano-oxidation could be a valuable technique for fundamental studies of two-level systems (TLS) in amorphous materials \cite{Lisenfeld2016decoherence} by enabling spatially targeted creation of TLS in acoustic \cite{maccabe2020nano} and superconducting \cite{chen2023phonon} devices. 

\bibliography{sample}

\vspace{1cm}

\noindent \textbf{Funding}\\
{This work was supported by the ARO/LPS Cross Quantum Technology Systems program (grant
W911NF-18-1-0103), the U.S. Department of Energy Office of Science National Quantum Information
Science Research Centers (Q-NEXT, award DE-AC02-06CH11357), the Institute for Quantum Information
and Matter, an NSF Physics Frontiers Center (grant PHY-1125565) with support of the Gordon and Betty
Moore Foundation, the Kavli Nanoscience Institute at Caltech, and the AWS Center for Quantum Computing. S.M. acknowledges support from the IQIM Postdoctoral Fellowship.}

\vspace{0.5cm}


\noindent \textbf{Disclosures.} The authors declare no conflicts of interest.

\noindent \textbf{Supplementary information.} is available in the online version of the paper. 

\noindent\textbf{Competing interests.} The authors declare no competing interests.

\noindent \textbf{Data Availability Statement.} Correspondence and requests for materials should be sent to OP (opainter@caltech.edu).

\newpage
\onecolumn
\section*{Supplementary Section}%

\subsection{\centering Experimental setup}%

\begin{figure*}[h!]
\includegraphics[width=\textwidth]{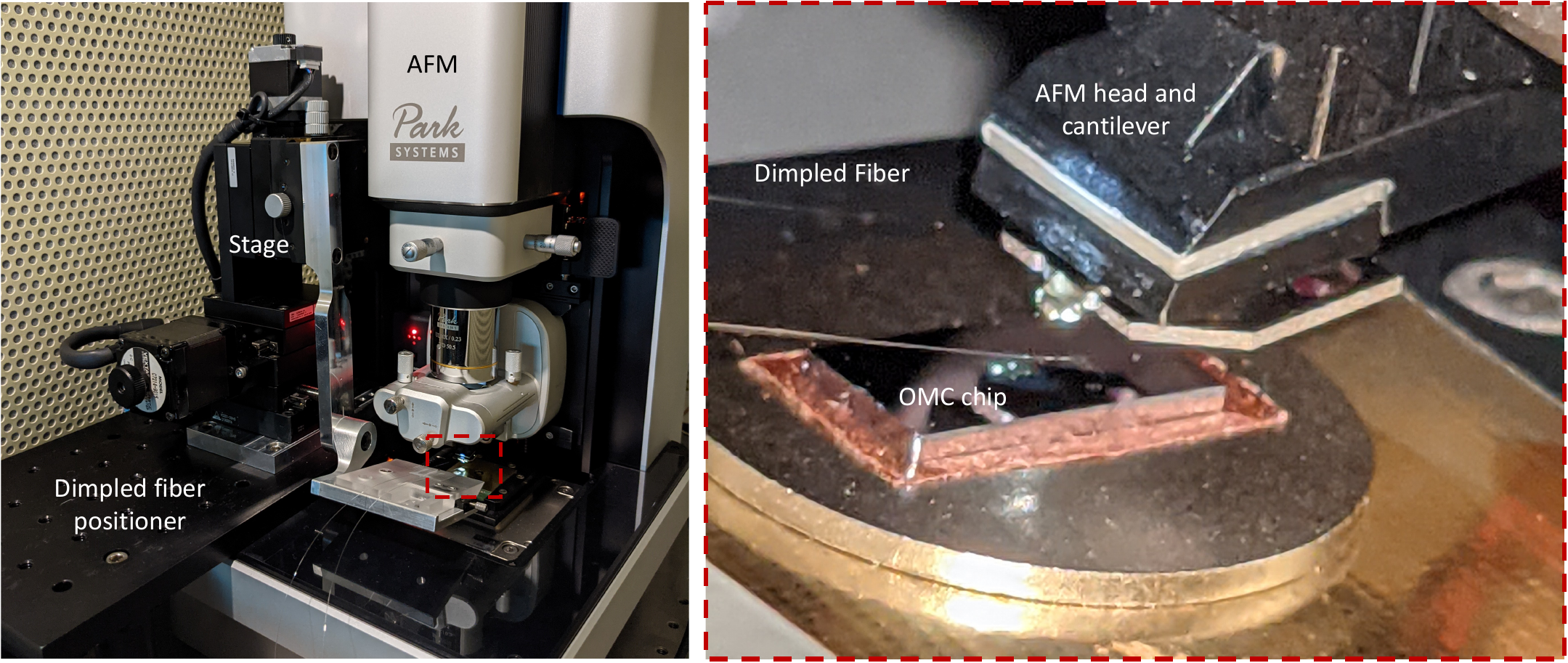}
\centering
\caption{\textbf{AFM integrated with optical test setup.} A dimpled fiber positioning stage is integrated into an AFM, allowing for real-time oxidation and optical testing. Image on the right shows a close-up view of the AFM cantilever, the dimpled fiber and the OMC chip.}
\label{fig:figs1}
\end{figure*}

To enable in-situ nano-oxidation and real-time tuning of OMC cavities, an optical test setup has been integrated into the enclosure of an AFM as shown in Fig. 5. The system has three sub-parts that are controlled independently: 1) the optical fiber positioning system, comprising a three-axes linear motorized positioning stage and two rotational positioning stages, 2) the sample mount of the AFM with x and y motional degrees of freedom, and 3) the AFM cantilever with z motional degree of freedom. Additionally, the AFM enclosure is equipped with a temperature and humidity sensor to ensure that environmental conditions are within the desired regime. The entire assembly sits on an optical table to minimize vibrations during the oxidation process.

\newpage
\subsection{\centering Characterization of nano-oxidation}%
\label{supplementary_characterization}
\begin{figure*}[h!] 
\includegraphics[width=\textwidth]{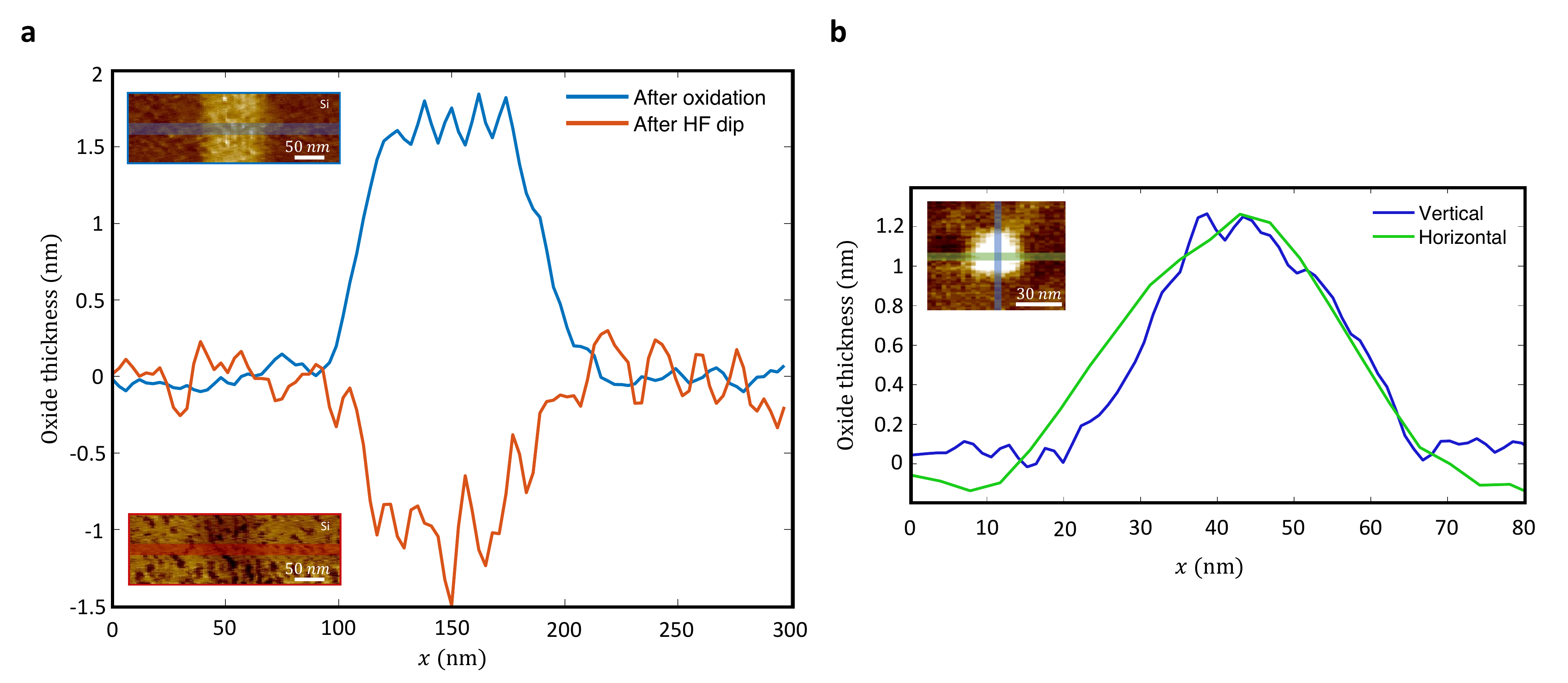} 
\centering
\caption{\textbf{Characterization of nano-oxidation on a silicon chip. a,} Height profile of the oxide slab created by nano-oxidation (blue trace), and the depth profile obtained after removal of the oxide slab using a hydrofluoric acid etch (red trace). Insets show the AFM scan of the oxide slab (top) and the etched trench (bottom) along with horizontal line cuts used to sample the traces on the main plot. \textbf{b,} Height profile of an oxide pixel along x and y axes indicating a lateral size of $\sim$25 nm. Inset shows the AFM scan of the oxide pixel along with x and y line cuts used to sample the traces on the main plot.}
\label{fig:figs2}
\end{figure*}

The nano-oxidation process generates oxide beneath and above the surface of silicon. Here we describe the procedure used to assess the ratio of oxide height and depth above and below the silicon surface, respectively. We first measure the post-oxidation height by performing AFM scans in the tapping mode. Subsequently, we etch the oxide in an anhydrous hydrofluoric acid vapor etching tool (SPTS PRImaxx uEtch), and perform AFM scans of the resulting trenches in silicon. The data obtained, as depicted in Figure \ref{fig:figs2}a, indicates that the oxide thickness beneath the surface is approximately 70$\%$ of the total measured thickness above the surface. Figure \ref{fig:figs2}b shows the characterization of a single oxide pixel generated in the mild oxidation mode. The size of the pixel is approximately 25nm along both x and y axes. We utilize such oxide pixels for fine-tuning of both optical and acoustic resonances in real-time as discussed in the main text.

\newpage
\subsection{\centering Effect of nano-oxidation on optical Q factor}%
\label{Effect_on_Q_factor}
\begin{figure*}[h!]
\includegraphics[width=12cm]{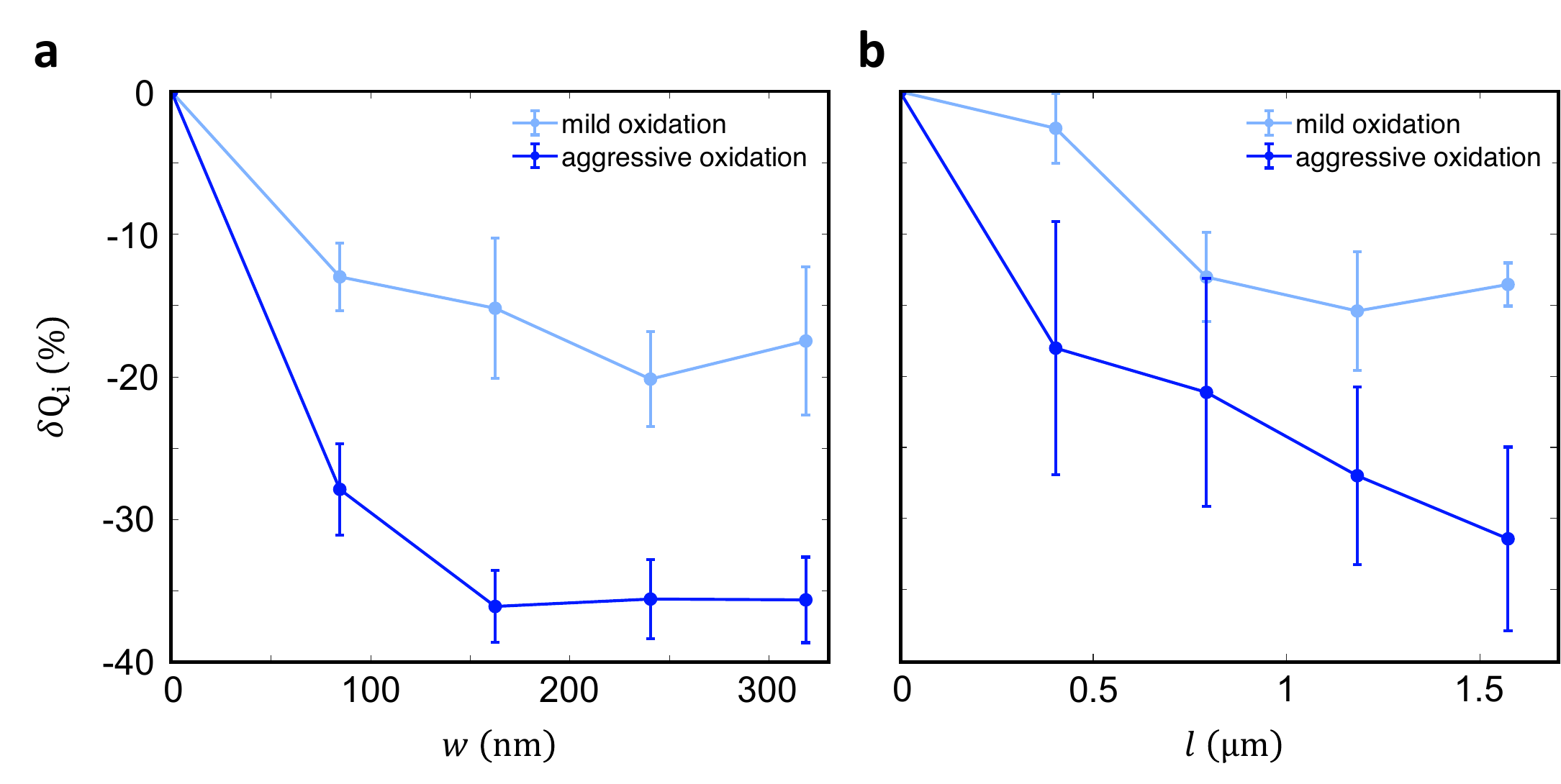}
\centering
\caption{\textbf{Effect of nano-oxidation on the optical Q factor during coarse tuning.}  Change in optical intrinsic Q-factor with respect to the \textbf{a}, width $w$ of the optics-focused oxidation pattern, and \textbf{b}, length $l$ of the acoustics-focused pattern. Data is obtained in both mild and aggressive oxidation modes. Error bars represent the standard deviation measured across five OMCs.}
\label{fig:figs3}
\end{figure*}

In addition to altering the mechanical and optical characteristics of the local silicon surface, the nano-oxidation process can introduce roughness and additional scattering sites. Consequently, we observe a reduction in the optical Q-factor during our coarse tuning experiments. As shown in Figure \ref{fig:figs3}, the reduction in Q-factor is more pronounced upon increasing the size of the oxidation patterns and the thickness of the oxide layer. For optics-focused patterns generated in regions with high electric field intensity, we observe a substantial impact on the optical Q-factor. A maximum reduction in $Q_i$ of approximately 35\% is noted for the aggressive oxidation case when $w<$ 300 nm. For the mild oxidation case, the reduction in $Q_i$ is less than 20\%. In the case of the acoustics-focused pattern, the reduction in $Q_i$ is less than 35\% for aggressive oxidation case, and less than 20\% for mild oxidation case when $l<$ 1.5 $\mu$m.

\newpage
\subsection{\centering FEM simulations for pixel-by-pixel oxidation and selectivity analysis}%
\label{Supplementary_D}
\begin{figure*}[h!]
\includegraphics[width=\linewidth]{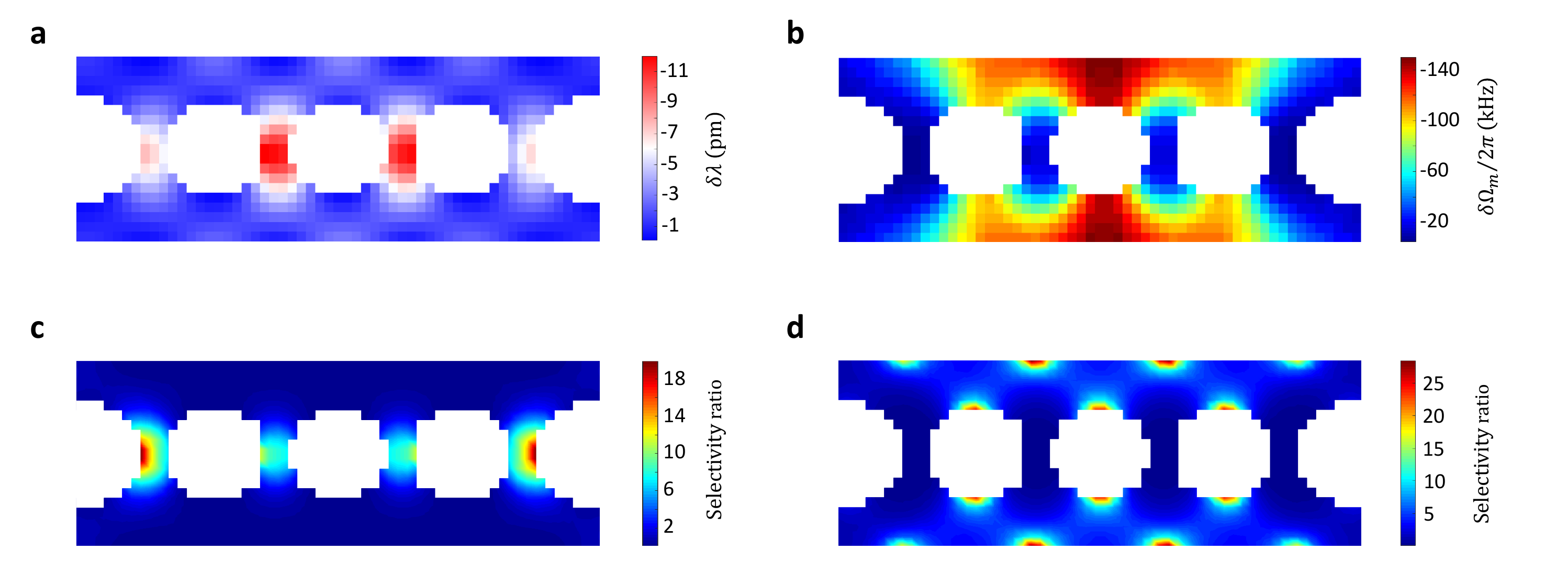}
\centering
\caption{\textbf{FEM simulations for pixel-by-pixel oxidation.} \textbf{a}, Optical resonance wavelength shift, $\delta\lambda$ and \textbf{b}, mechanical resonance frequency shift, $\delta\Omega/2\pi$ simulated as a function of the location of the oxide pixel. Each oxide pixel is 25x25 nm in size with a thickness of 3.2 nm. \textbf{c}, Interpolated normalized optical selectivity, $\delta\lambda/\delta\Omega$ and \textbf{d}, interpolated normalized acoustic selectivity, $\delta\Omega/\delta\lambda$. The normalization is performed with respect to the maximum shifts obtained in the pixel-by-pixel oxidation profiles in panels a and b.}
\label{fig:figs7}
\end{figure*}

Figure \ref{fig:figs7}a and b show the results of FEM simulations where we mapped out how the location of a single oxidation pixel affects the optical wavelength and acoustic frequency, respectively. These profiles closely match the optical intensity and acoustic displacement profiles shown in Fig. \ref{fig:fig1}. This is expected because oxidation at points with larger optical field concentration leads to larger change in optical energy, and thereby in the optical resonance wavelength. Similarly, oxidation at the points of maximum displacement leads to a larger fractional change in effective mass of the acoustic mode, and thereby in the acoustic resonance frequency. The exact material properties of the thin film oxide such as the density, Young's modulus and relative permittivity are specific to the oxidation process. In our approach, we use these material properties as variables and then we find what scaling allows us to match the coarse tuning experiment (Fig.~\ref{fig:fig2}) and simulations. This way, we found that the density, Young's modulus and relative permittivity of oxide are 3500 kg/$\text{m}^\text{3}$, 70 GPa and 2, respectively. In the acoustic simulations, we use the anisotropic elasticity tensor of silicon with ($C_{11} , C_{12} , C_{44}$) = (166, 64, 80) GPa and assume a [110] crystallographic orientation for the x-axis.

In Figure \ref{fig:figs7}c and d, we plot the normalized optical selectivity, $\delta\lambda/\delta\Omega$ and normalized acoustic selectivity,  $\delta\Omega/\delta\lambda$. Here, $\delta\lambda$ and $\delta\Omega$ are normalized with respect to the maximum value calculated in the pixel-by-pixel oxidation profiles in Figure \ref{fig:figs7}a, b. In the case of fine tuning experiments where we seek small frequency shifts, we concentrate on the high selectivity regions. However, the locations with maximum optical selectivity, $\delta\lambda/\delta\Omega$ do not coincide with those that produce maximum optical tuning $\delta\lambda$. Likewise, the regions with maximum acoustic selectivity are relatively small and localized at the edge of the silicon-air boundary. As a result, for coarse tuning experiments, it is more practical to use large oxide patches as opposed to individual pixels to achieve large tuning at the expense of selectivity. Details of the pattern generation algorithm used to generate the large oxide patches are given in Appendix E.

\newpage
\subsection{\centering Nano-oxidation inverse pattern generation algorithm.}%
\label{supplement:inverse_algo}
\begin{figure*}[h!]
\includegraphics[width=\textwidth]{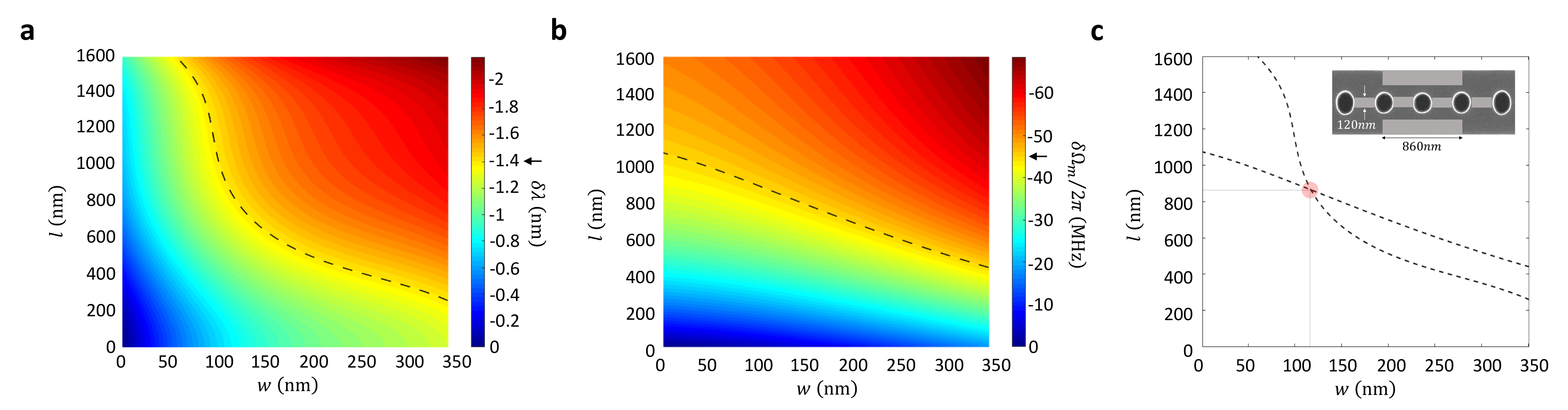}
\centering
\caption{\textbf{Pattern generation matrices for coarse tuning.} An example pair of experimentally determined pattern generation matrices used for aggressive optical - aggressive acoustic oxidation. \textbf{a}, Optical wavelength shift, $\delta\lambda$ and \textbf{b}, mechanical frequency shift, $\delta\Omega_m/2\pi$ as a function of the width of the optics-focused component, $w$, and length of the acoustics-focused component, $l$ of a joint pattern. All other dimensions and parameters are kept constant. The matrices are generated by spline interpolation of experimentally measured frequency shift data. \textbf{c}, Pattern dimension contours corresponding to $\delta\lambda=-1.4$ nm and $\delta\Omega_m/2\pi=-45$ MHz. The intersection of both curves is used to determine $l$ and $w$ for the joint pattern used for coarse tuning. Inset shows the joint pattern corresponding to the calculated dimensions.}
\label{fig:figs4}
\end{figure*}

The results from coarse tuning experiments in Fig.~\ref{fig:fig2} of the main text show frequency shifts generated by either the optics-focused or the acoustic-focused pattern type. In practice, we require a union of both types of patterns to tune both optical and acoustic resonances. To enable the generation of such joint patterns, we first interpolate results from coarse oxidation experiments to create an oxidation map with frequency shifts measured with just the optics-focused and acoustic-focused patterns. Our inverse pattern generation algorithm uses this map to find a suitable joint pattern to target the desired simultaneous shift in optical and acoustic resonances. The pattern search begins with mild oxidation patterns and progresses to aggressive oxidation patterns if larger frequency shifts are desired. A pattern is deemed suitable if the calculated resonance shifts can target the desired optical wavelength to within 200 pm and the desired acoustic resonance to within 5 MHz. The pattern generation maps presented in Fig.~\ref{fig:figs4}a,b correspond to the aggressive optical - aggressive acoustic oxidation scenario. There are three other pairs of pattern generation maps corresponding using mild optical - mild acoustic, mild optical - aggressive acoustic, and mild optical - aggressive acoustic oxidation scenarios. To illustrate the pattern search process, Figs.~\ref{fig:figs4}a, b show contours corresponding to an example scenario where shifts, $\delta\lambda=-1.4$ nm and $\delta\Omega_m/2\pi=-45$ MHz, are desired. As shown in Fig.~\ref{fig:figs4}c, the intersection of these two contours allows us to estimate the required joint pattern with a width, $w\sim$ 120nm for the optics-focused component and length, $l\sim$ 860nm for the acoustics-focused component.

\newpage
\subsection{\centering Simultaneous tuning of three OMCs: step-by-step process}%
\label{3_OMC_tuning_supplement}
\begin{figure*}[h]
\includegraphics[width=\textwidth]{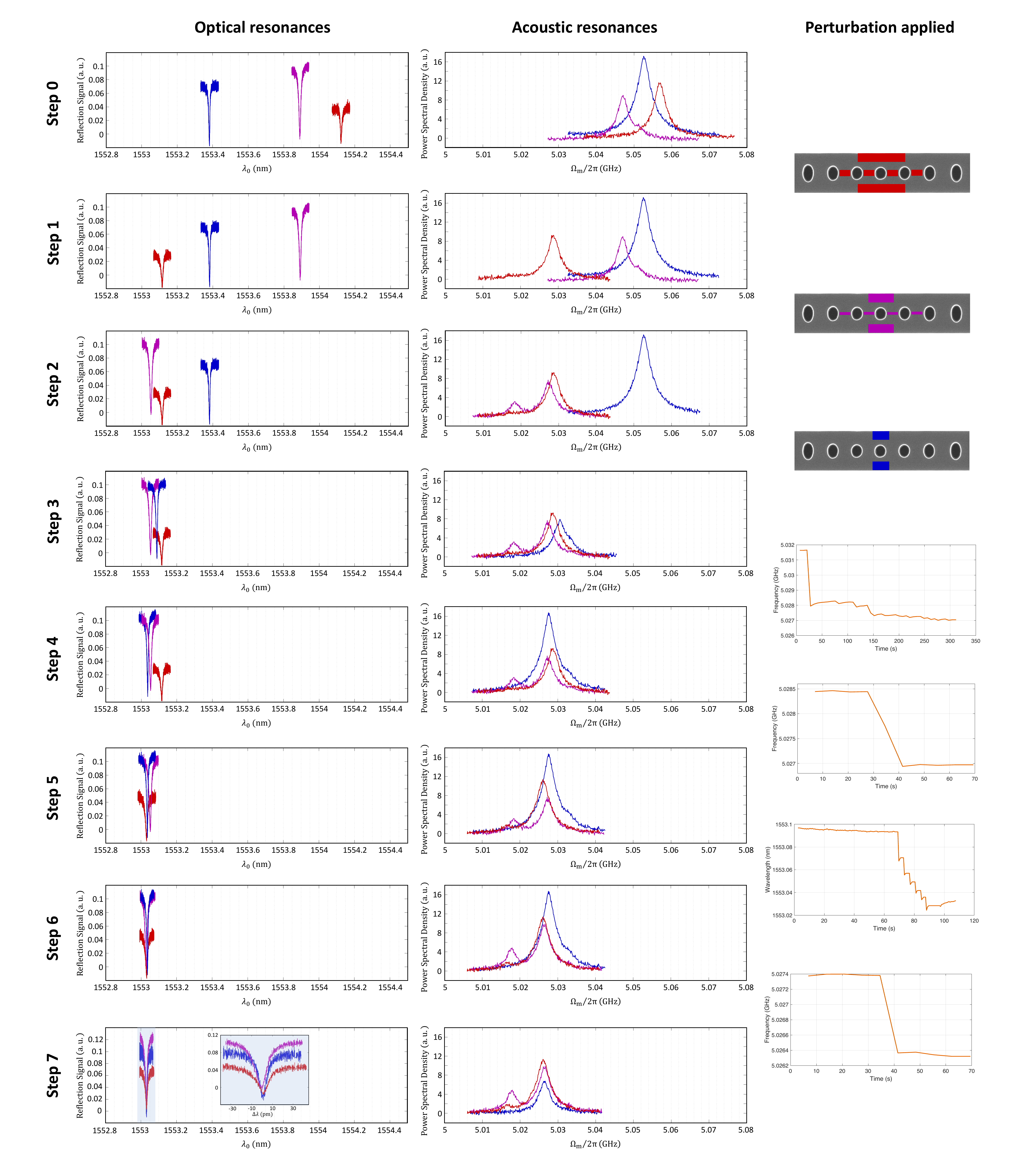}
\centering
\caption{\textbf{Simultaneous optical and acoustic frequency alignment of three OMCs.} Step 0 shows the initial condition of three OMCs. In steps 1-3, we generate coarse tuning patterns via the inverse pattern algorithm and apply them to the corresponding OMCs. These patterns are color coded for each OMC and shown in the right column of the figure. In steps 4-7, we perform real-time fine tuning of both optical and acoustic resonances. Optical and acoustic spectra monitored during the nano-oxidation sequence in real-time are shown in the right column of the figure for each step. In step 7, the final states of e optical and acoustic resonances on the three devices are shown. The inset shows a zoomed-in plot of the three optical resonances. This seven step process allows for optical and acoustic resonance matching to within 2 pm and 500 kHz, respectively.}
\label{fig:figs5}
\end{figure*}

\noindent In this section, we describe in detail the step-by-step tuning protocol used for frequency alignment of three OMC cavities. This involves a series of coarse tuning steps followed by real-time fine tuning steps. The experimental results for each step are shown in Fig.\,\ref{fig:figs5}.

\begin{enumerate}
\item  \textbf{Step 0}: We started with three OMC cavities (red, magenta, blue) whose optical wavelengths and acoustic frequencies are $\lambda$ = (1553.891, 1553.381, 1553.123) pm and $\Omega_m/2\pi$ = (5.057, 5.047, 5.053) MHz respectively.

\item \textbf{Step 1-3}: We feed  $\lambda$ and $\Omega_m/2\pi$ to the inverse pattern generation algorithm which generates patterns to tune each OMC to the target frequencies. 

\item \textbf{Step 4-7}: At the end of step 3, all 3 OMC cavities are within a suitable range for fine tuning (typically 200 pm and 5 MHz for optics and acoustic resonances). We then employ pixel-by-pixel oxidation to fine tune the resonances while monitoring their values in real-time. The corresponding real-time wavelengths and frequencies are shown in the right most column of Fig.\,\ref{fig:figs5}.
\end{enumerate}

\noindent At the end of three coarse and four fine tuning steps, we achieve optical wavelengths of 1553.031 nm, 1553.029 nm and 1553.031 nm and acoustic frequencies of 5.0261 GHz, 5.0264 GHz and 5.0266 GHz for the three OMCs. This indicates simultaneous optical wavelength and acoustic frequency matching within 2 pm and 500 kHz, respectively.

\subsection{\centering Frequency distribution of fabricated devices}%

\begin{figure*}[h]
\includegraphics[width=14cm]{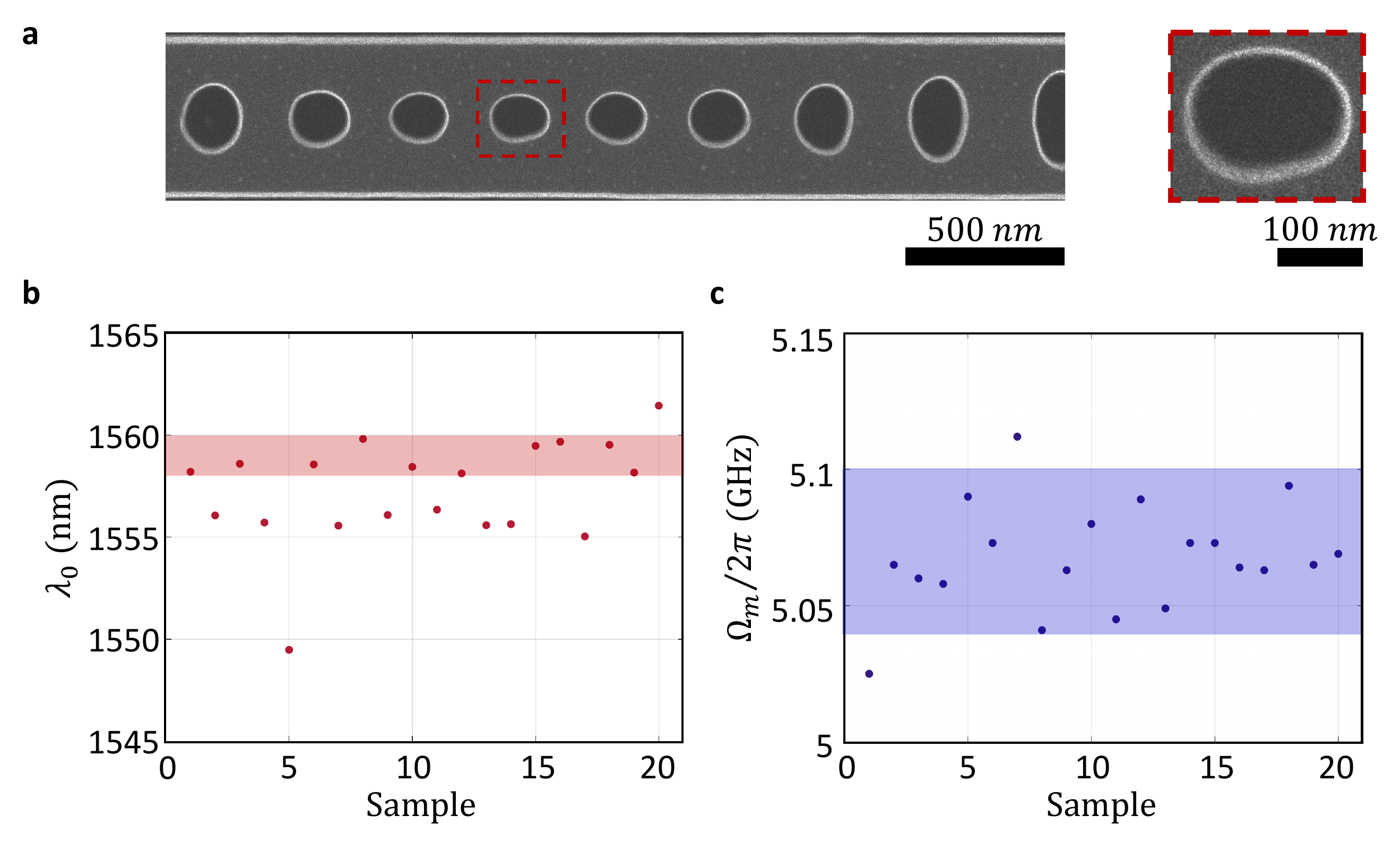}
\centering
\caption{\textbf{Resonance frequency distribution of 20 OMCs fabricated with nominally identical device geometry.} \textbf{a}, SEM image of the defect unit cells of a randomly chosen OMC cavity. Inset on the right shows the magnified view of an ellipse with irregular boundaries. \textbf{b},  Optical resonance wavelengths and \textbf{c}, acoustic resonance frequencies of 20 OMC cavities.  Shaded areas highlight cavities that can be independently tuned to the same optical wavelength (red) or acoustic frequency (blue) with maximum tuning ranges set at 2 nm and 60 MHz, respectively. The data indicates that approximately 50\% (90\%) of the devices can be frequency-matched for only optics (acoustics).}
\label{fig:figs6}
\end{figure*}

The OMC cavity design in this study incorporates geometric features with critical dimensions smaller than 100 nm. In our fabrication process, electron beam lithography and reactive ion etching offer a feature size precision of approximately 5 nm \cite{maccabe2020nano}. In Figure \ref{fig:figs6}a, a scanning electron microscope (SEM) image of a randomly selected OMC highlights geometric imperfections in the fabricated device. To assess how these imperfections manifest in the measured optical and acoustic resonances, we fabricated 20 devices with nominally identical geometry positioned adjacent to each other on a chip.

Figures \ref{fig:figs6}b and c show the measured optical wavelengths with a standard deviation of 2.58 nm and acoustic resonance frequencies with a standard deviation of 19.8 MHz, respectively. The highlighted area in Figures \ref{fig:figs6}b (c), with a tuning range of 2 nm (60 MHz), signifies the number of devices that can be brought into resonance solely through optical (acoustic) tuning. The data indicates that approximately 50\% (90\%) of OMCs fabricated with the identical geometric design can be frequency-matched optically (acoustically) after fabrication.

\end{document}